\documentclass[american,aps,prl,superscriptaddress,reprint]{revtex4-1}
\usepackage[T1]{fontenc}
\usepackage[utf8]{inputenc}
\usepackage{color}
\usepackage{babel}
\usepackage{mathtools}
\usepackage{amsmath}
\usepackage{graphicx}
\usepackage[unicode=true,
 bookmarks=true,bookmarksnumbered=false,bookmarksopen=false,
 breaklinks=false,pdfborder={0 0 0},pdfborderstyle={},backref=false,colorlinks=true]
 {hyperref}
\hypersetup{
 pdfborderstyle={},pdfborderstyle={},allcolors=blue}

\makeatletter

\newcommand*\LyXThinSpace{\,\hspace{0pt}}

\usepackage{babel}
\usepackage{babel}

\usepackage{bbold}

\usepackage{algpseudocode}

\makeatother

\begin{document}
\title{Experimental characterization of a spin quantum heat engine }
\author{John P. S. Peterson }
\thanks{These authors contributed equally to this work.}
\affiliation{Institute for Quantum Computing and Department of Physics and Astronomy,
University of Waterloo, Waterloo N2L 3G1, Ontario, Canada}
\author{Tiago B. Batalhão}
\thanks{These authors contributed equally to this work.}
\affiliation{Centro de Ciências Naturais e Humanas, Universidade Federal do ABC,
Avenida dos Estados 5001, 09210-580 Santo André, São Paulo, Brazil}
\affiliation{Singapore University of Technology and Design, 8 Somapah Road, Singapore
487372, Singapore }
\affiliation{Centre for Quantum Technologies, National University of Singapore,
3 Science Drive 2, Singapore 117543, Singapore}
\author{Marcela Herrera}
\thanks{These authors contributed equally to this work.}
\affiliation{Centro de Ciências Naturais e Humanas, Universidade Federal do ABC,
Avenida dos Estados 5001, 09210-580 Santo André, São Paulo, Brazil}
\author{Alexandre M. Souza}
\affiliation{Centro Brasileiro de Pesquisas Físicas, Rua Dr. Xavier Sigaud 150,
22290-180 Rio de Janeiro, Rio de Janeiro, Brazil}
\author{Roberto S. Sarthour}
\affiliation{Centro Brasileiro de Pesquisas Físicas, Rua Dr. Xavier Sigaud 150,
22290-180 Rio de Janeiro, Rio de Janeiro, Brazil}
\author{Ivan S. Oliveira}
\affiliation{Centro Brasileiro de Pesquisas Físicas, Rua Dr. Xavier Sigaud 150,
22290-180 Rio de Janeiro, Rio de Janeiro, Brazil}
\author{Roberto M. Serra}
\email{Electronic Address: serra@ufabc.edu.br}

\affiliation{Centro de Ciências Naturais e Humanas, Universidade Federal do ABC,
Avenida dos Estados 5001, 09210-580 Santo André, São Paulo, Brazil}
\begin{abstract}
Developments in the thermodynamics of small quantum systems envisage
non-classical thermal machines. In this scenario, energy fluctuations
play a relevant role in the description of irreversibility. We experimentally
implement a quantum heat engine based on a spin-1/2 system and nuclear
magnetic resonance techniques. Irreversibility at microscope scale
is fully characterized by the assessment of energy fluctuations associated
with the work and heat flows. We also investigate the efficiency lag
related to the entropy production at finite time. The implemented
heat engine operates in a regime where both thermal and quantum fluctuations
(associated with transitions among the instantaneous energy eigenstates)
are relevant to its description. Performing a quantum Otto cycle at
maximum power, the proof-of-concept quantum heat engine is able to
reach an efficiency for work extraction ($\eta\approx42$\%) very
close to its thermodynamic limit ($\eta=44$\%). 
\end{abstract}
\maketitle
Quantum thermal machines perform a thermodynamic cycle employing quantum
systems as the working medium. This notion was introduced long ago
when Scovil and Schulz-Dubois recognized a three-level maser as a
kind of heat engine \cite{Scovil1959}, and since then many theoretical
proposals for thermodynamical cycles at the quantum scale have been
discussed \cite{Alicki1979,Geva1992,Kieu2004,Quan2007,Haenggi2009,Linden2010,Correa2014,Palao2015,Uzdin2014,Uzdin2015,Binder2015,Gelbwaser-Klimovsky2013,Scully2003,Brunner2012,Gallego2014,Kosloff2014,Giovannetti2015,Pleno2015,Alecce2015,Hofer2016,Campisi2016,Goold2016,Vinjanampathy2016,Elouard2017,Dechant2015,watanabe2017,Reid2017,Feldmann2003,Plastina2014,Shiraishi2017,Benenti2017}.
Microscopic quantum heat engines may operate at a scale where both
thermal and quantum fluctuations are relevant. The thermodynamic description
of such devices operating at finite time should also include the inherent
non-deterministic nature of the quantum evolution and non-equilibrium
features. In this context, quantities as work, heat, and entropy production
are associated with statistical distributions that satisfy fluctuation
theorems \cite{Esposito2009,Campisi2011,Hanggi2015} for a thermodynamical
cycle \cite{Campisi2014,Campisi2015}.

The enthusiastic interest in quantum thermal machines has grown with
the possibility to control non-equilibrium dynamics of microscopic
systems, achievable in platforms such as: trapped ions \cite{Abah2012,Rosnagel2014},
quantum dots \cite{Kennes2013,Sanchez2013,Sothmann2014}, single electron
boxes \cite{Berg2015}, optomechanical oscillators \cite{Zhang2014,Bergenfeldt2014,Elouard2015,Brunelli2015},
etc. Some experimental success related to the implementation of micro-scale
heat engines have been reported in a context where quantum coherence
effects are not prominent (which can be regarded as a classical context)
\cite{Hugel2002,Steeneken2011,Blickle2011,Brantut2013,Thierschmann2015,Schmidt2018}.
Recently, a single trapped ion was employed as a working medium to
perform a thermodynamic cycle \cite{Rossnagel2016}. Despite this
latter implementation being based on a single quantum system, the
operating temperatures are such that the thermal energy is considerably
higher than the energy level separation of the magnetic trap. As a
consequence, effects of quantum fluctuations are dwarfed by thermal
fluctuations allowing a classical description.

The full characterization of a finite-time operation of a quantum
heat engine may also be associated with the assessment of the probability
distribution of energy fluctuations, that can take the form of work
or heat flow \cite{Holubec2017}. This assessment embodies significant
experimental challenges that remained elusive up to now.

In the present contribution, we used a Nuclear Magnetic
Resonance (NMR) setup \cite{Oliveira2007} to implement and characterize
a quantum version of the Otto cycle \cite{Kieu2004}. As a proof-of-concept
implementation of a quantum heat engine operating at finite time,
we employed a $^{13}$C-labeled CHCl$_{3}$ liquid sample diluted
in Acetone-D6 and a 500 MHz Varian NMR spectrometer. The spin 1/2
of the $^{13}$C nucleus is the working medium whereas the $^{1}$H
nuclear spin will be used as a heat bus. High radio-frequency (rf)
modes near to Hydrogen Larmor frequency plays the role of the hot
environment while low rf modes near to Carbon resonance frequency
plays the role of the cold environment. Chlorine isotopes' nuclei
provide mild environmental effects. An interferometric method \cite{Dorner2013,Mazzola2013,Batalhao2014,Goold2014,Batalhao2015}
is applied to assess energy fluctuations to characterize the work
and heat statistics as well as the irreversibility aspects of this
spin engine. The operation regime is such that the typical thermal
energy scale is of the same order of the typical separation of the
quantum energy levels, turning the effects of quantum fluctuations
as important as the ones from thermal fluctuations. We have also experimentally
endorsed an expression for the efficiency lag related to the entropy
production that hinders the implemented engine to attain the Carnot
efficiency at finite time. The cycle was established at different
finite-time regimes, ranging from a very irreversible to one with
almost maximum efficiency, allowing the identification of the maximum
power operation.

\begin{figure}
\begin{centering}
\includegraphics[width=0.98\columnwidth]{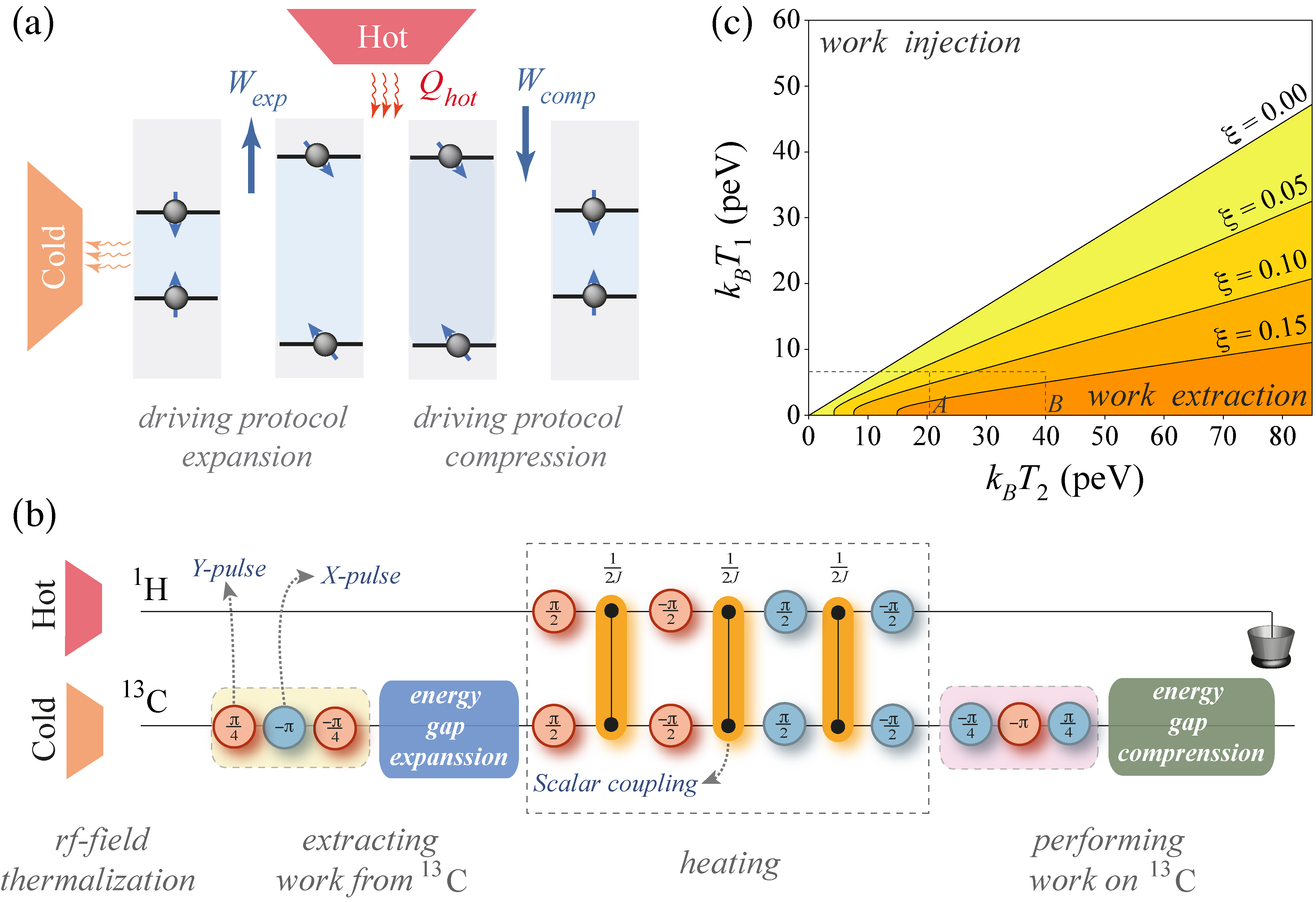} 
\par\end{centering}
\caption{Quantum heat engine schematics. (a) Thermodynamic cycle employing
a spin 1/2 as working medium. (b) Simplified pulse sequence of the
experimental protocol. $^{1}$H and $^{13}$C nuclear spins are initially
prepared in thermal states corresponding to hot and cold spin temperatures,
respectively. Blue (red) circles represent $x$ ($y$) rotations by
the displayed angle produced by transverse rf pulses. Orange connections
stand for free evolutions under the scalar interaction ($\mathcal{H}_{J}$)
during the time displayed above the symbol. The unitary driving for
the energy gap expansion (compression) protocol is implemented by
a time-modulated rf field resonant with the $^{13}$C nuclear spin. 
The Hydrogen nucleus is used to deliver the heat
at the proper part of the cycle, working as a heat bus. (c) Required
temperatures for work extraction at finite-time operation mode. The
engine extracts work only if the hot ($T_{2}$) and cold ($T_{1}$)
source temperatures correspond to a point below the curve defined
by the energy level transition probability $\xi$.}
\label{fig:schematics-1} 
\end{figure}

The quantum version of the Otto cycle \cite{Kieu2004,Alecce2015}
consists of a four-stroke protocol as illustrated in Fig.~\ref{fig:schematics-1}(a).

\emph{Cooling stroke. }Using spatial average techniques employed by
rf and gradient fields, the $^{13}\text{C}$ nuclear spin is initially
chilled to a pseudo-thermal state, equivalent to $\rho_{0}^{eq,1}=\left.e^{-\beta_{1}\mathcal{H}_{1}^{\text{C}}}\right/Z_{1}$
\cite{Micadei2017,SuppMat}, at a cold inverse spin temperature $\beta_{1}=\left(k_{B}T_{1}\right)^{-1}$,
where $Z_{1}=\text{tr}\left(e^{-\beta_{1}\mathcal{H}_{1}^{\text{C}}}\right)$
is the partition function, $k_{B}$ is the Boltzmann constant, $T_{1}$
is the absolute spin temperature of the cold reference state, and
the Hamiltonian $\mathcal{H}_{1}^{\text{C}}$ will be defined latter.

\emph{Expansion stroke.} The working medium Hamiltonian is driven
by a time-modulated rf field resonant with the $^{13}$C nuclear spin.
Initially it can be described by $\mathcal{H}_{1}^{\text{C}}=-h\nu_{1}\sigma_{y}^{\text{C}}/2$
(with the rf-field intensity adjusted such that $\nu_{1}=2.0$~kHz
and $\sigma_{x,y,z}^{\text{C}}$ being the Pauli spin operators for
$^{13}$C nuclear spin), in a rotating frame at the $^{13}$C Larmor
frequency ($\text{\ensuremath{\approx}125}$~MHz). From $t=0$ up
to $t=\tau$, the system Hamiltonian is driving according to $\mathcal{H}_{exp}^{\text{C}}\left(t\right)=-\frac{1}{2}h\nu(t)\left(\cos\left(\frac{\pi t}{2\tau}\right)\sigma_{x}^{\text{C}}+\sin\left(\frac{\pi t}{2\tau}\right)\sigma_{y}^{\text{C}}\right)$,
expanding ($exp$) the nuclear spin energy gap linearly as $\nu(t)=\nu_{1}\left(1-\frac{t}{\tau}\right)+\nu_{2}\frac{t}{\tau}$
(with $\nu_{2}=3.6$~kHz and $t\in\left[0,\tau\right]$). The energy
gap expansion happens in a driving time length, $\tau$, that will
be varied in different experiments between $100\ensuremath{~}\mu$s
and $700\ensuremath{~}\mu$s. The driving time length ($\propto10^{-4}$s)
is much shorter than the typical decoherence time scales, which are
on the order of seconds. In this way, we can describe this process
by a unitary evolution, $\mathcal{U}_{\tau}$ \cite{Batalhao2014,Batalhao2015,SuppMat},
that drives the $^{13}$C nuclear spin to an out-of-equilibrium state
($\rho_{\tau}^{\text{C}}$), which is, in general, not diagonal in
the energy eigenbasis of the final Hamiltonian of the expansion protocol,
$\mathcal{H}_{2}^{\text{C}}=\mathcal{H}_{exp}^{\text{C}}\left(\tau\right)=-h\nu_{2}\sigma_{x}^{\text{C}}/2$.

\emph{Heating stroke}. The working medium ($^{13}$C nucleus) exchanges
heat with the $^{1}$H nuclear spin, which was initially prepared
in a higher temperature \cite{Micadei2017,SuppMat} than the $^{13}$C
nuclear spin, reaching full thermalization at the hot inverse spin
temperature $\beta_{2}=\left(k_{B}T_{2}\right)^{-1}$. The full thermalization
process is effectively implemented by a sequence of free evolutions
under the scalar interaction, $\mathcal{H}_{J}=\frac{1}{2}hJ\sigma_{z}^{\text{H}}\sigma_{z}^{\text{C}}$
(with $J\approx215.1$~Hz), between both nuclei and rf pulses to
produce suitable rotations as sketched in Fig.~\ref{fig:schematics-1}(b).
After thermalization, the state of the $^{13}\text{C}$ nuclei is
the hot equilibrium state, equivalent to $\rho_{0}^{\text{eq},2}=\left.e^{-\beta_{2}\mathcal{H}_{2}^{\text{C}}}\right/Z_{2}$.

\emph{Compression stroke. }Subsequently, an energy gap compression
is performed, according to the time-reversed process \cite{Camati2018}
of the expansion protocol, i.e., the Hamiltonian is driven in a way
that $\mathcal{H}_{comp}^{\text{C}}\left(t\right)=-\mathcal{H}_{exp}^{\text{C}}\left(\tau-t\right)$.

Many cycles of this proof-of-concept experiment can be performed repeating
successively the pulse sequence protocol described in Fig.~\ref{fig:schematics-1}(b).
It is interesting to note that each experimental run involves spatial
averages on a diluted liquid sample containing about $10^{17}$ molecules,
which can be regarded as noninteracting with each other due to the
sample dilution. Each experimental result for the quantities of interest
represents an average over many copies of a single molecular spin
engine.

The finite-time (expansion and compression) driven processes are associated
with transitions among the instantaneous eigenstates of the working
medium Hamiltonian (see Fig.~S3 of \cite{SuppMat}) resulting in
entropy production \cite{Batalhao2015,Camati2016}, which is the main
source of irreversibility in the implemented cycle. In this way, quantum
coherence also contributes to the irreversibility \cite{Brandner2016,Brandner2017,Camati2019}.

Considering the aforementioned description of the finite-time thermodynamical
cycle, we can write the average values of the extracted work ($W_{eng}$)
from the engine and the absorbed heat ($Q_{hot}$) from the $^{1}$H
nuclear spin as 
\begin{align}
\left\langle W_{eng}\right\rangle  & =\frac{h}{2}\left(\nu_{2}-\nu_{1}\right)\left[\tanh\left(\beta_{1}h\nu_{1}\right)-\tanh\left(\beta_{2}h\nu_{2}\right)\right]\nonumber \\
 & -h\xi\left[\nu_{1}\tanh\left(\beta_{2}h\nu_{2}\right)+\nu_{2}\tanh\left(\beta_{1}h\nu_{1}\right)\right]\;,\label{meanwork}
\end{align}
\begin{align}
\left\langle Q_{hot}\right\rangle  & =\frac{h}{2}\nu_{2}\left[\tanh\left(\beta_{1}h\nu_{1}\right)-\tanh\left(\beta_{2}h\nu_{2}\right)\right]\nonumber \\
 & -\xi h\nu_{2}\tanh\left(\beta_{1}h\nu_{1}\right)\;,\label{heat}
\end{align}
where $\xi=\left|\left\langle \Psi_{\pm}^{2}\mid\mathcal{U}_{\tau}\mid\Psi_{\mp}^{1}\right\rangle \right|^{2}=\left|\left\langle \Psi_{\pm}^{1}\mid\mathcal{V}_{\tau}\mid\Psi_{\mp}^{2}\right\rangle \right|^{2}$
are the transition probabilities between the instantaneous eigenstates
$\left|\Psi_{\pm}^{1}\right\rangle $ ($\left|\Psi_{\pm}^{2}\right\rangle $)
of the Hamiltonian $\mathcal{H}_{1}^{\text{C}}$ ($\mathcal{H}_{2}^{\text{C}}$)
and $\mathcal{V}_{\tau}$ is the unitary evolution describing the
compression protocol, satisfying $\mathcal{V}_{\tau}=\mathcal{U}_{\tau}^{\dagger}$.
The nuclear spin system operates as a heat engine when $\langle W_{eng}\rangle>0$,
otherwise work is being injected in the device during the cycle. Two
conditions must be met to allow work extraction. The first is the
requirement that $\left(\nu_{2}-\nu_{1}\right)\left[\tanh\left(\beta_{1}h\nu_{1}\right)-\tanh\left(\beta_{2}h\nu_{2}\right)\right]\geq0$,
which is equivalent to the classical-scenario bound, $1\leq\nu_{2}/\nu_{1}\leq T_{2}/T_{1}$.
The second condition imposes a limit on the admitted transition probability
among the energy levels, which reads 
\begin{align}
\xi & \leq\frac{\left(\nu_{2}-\nu_{1}\right)\left[\tanh\left(\beta_{1}h\nu_{1}\right)-\tanh\left(\beta_{2}h\nu_{2}\right)\right]}{2\left[\nu_{1}\tanh\left(\beta_{2}h\nu_{2}\right)+\nu_{2}\tanh\left(\beta_{1}h\nu_{1}\right)\right]}.\label{transprob}
\end{align}
This condition, illustrated in Fig.~\ref{fig:schematics-1}(c), is
related to the rapidity of the energy gap expansion (compression)
protocol and to the fact that the driving Hamiltonian does not commute
at different times. For a given protocol (that sets the $\xi$ value)
the condition \eqref{transprob} only depends on the energy gap compression
factor, $r=\nu_{2}/\nu_{1}$ ($r\simeq1.8$ in our experiment). The
system operates in the working extraction mode if the point that characterizes
the temperature of both heat sources lies below the contour curve
in Fig.~\ref{fig:schematics-1}(c) for a given transition probability.

The spin-engine efficiency can be written also in terms of the energy
level transition probability as 
\begin{align}
\eta & =\frac{\left\langle W_{eng}\right\rangle }{\left\langle Q_{hot}\right\rangle }=1-\frac{\nu_{1}}{\nu_{2}}\frac{\left(1-2\xi\mathcal{F}\right)}{\left(1-2\xi\mathcal{G}\right)},\label{efficiency1}
\end{align}
where $\mathcal{F}=\tanh\left(\beta_{2}h\nu_{2}\right)\left(\tanh\left(\beta_{2}h\nu_{2}\right)-\tanh\left(\beta_{1}h\nu_{1}\right)\right)^{-1}$
and $\mathcal{G}=\left.\mathcal{F}\tanh\left(\beta_{1}h\nu_{1}\right)\right/\tanh\left(\beta_{2}h\nu_{2}\right)$.
The Otto limit ($\eta_{\text{Otto}}$) is recovered in an adiabatic
(transitionless, i.e. $\xi=0$) driving. On the other hand, in the
finite-time regime the efficiency \eqref{efficiency1} decreases as
$\xi$ increases. Alternatively, we can derive an expression for the
engine efficiency in terms of efficiency lags (associated with entropy
production \cite{Feldmann2003,Plastina2014,Shiraishi2017,Batalhao2015})
as $\eta=\eta_{\text{Carnot}}-\mathcal{L}$, and the lag is given
by \cite{SuppMat} 
\begin{align}
\mathcal{L} & =\frac{\mathcal{S}\left(\left.\mathcal{U}_{\tau}\rho_{0}^{eq,1}\mathcal{U}_{\tau}^{\dagger}\right\Vert \rho_{0}^{eq,2}\right)+\mathcal{S}\left(\left.\mathcal{V}_{\tau}\rho_{0}^{eq,2}\mathcal{V}_{\tau}^{\dagger}\right\Vert \rho_{0}^{eq,1}\right)}{\beta_{1}\left\langle Q_{hot}\right\rangle },\label{efficiency2}
\end{align}
where $\mathcal{S}\left(\left.\rho_{a}\right\Vert \rho_{b}\right)=\text{tr}\left[\rho_{a}\left(\ln\rho_{a}-\ln\rho_{b}\right)\right]$
is the relative entropy and $\eta_{\text{Carnot}}=1-\left.T_{1}\right/T_{2}$
the standard Carnot efficiency.

\begin{figure}
\begin{centering}
\includegraphics[width=0.99\columnwidth]{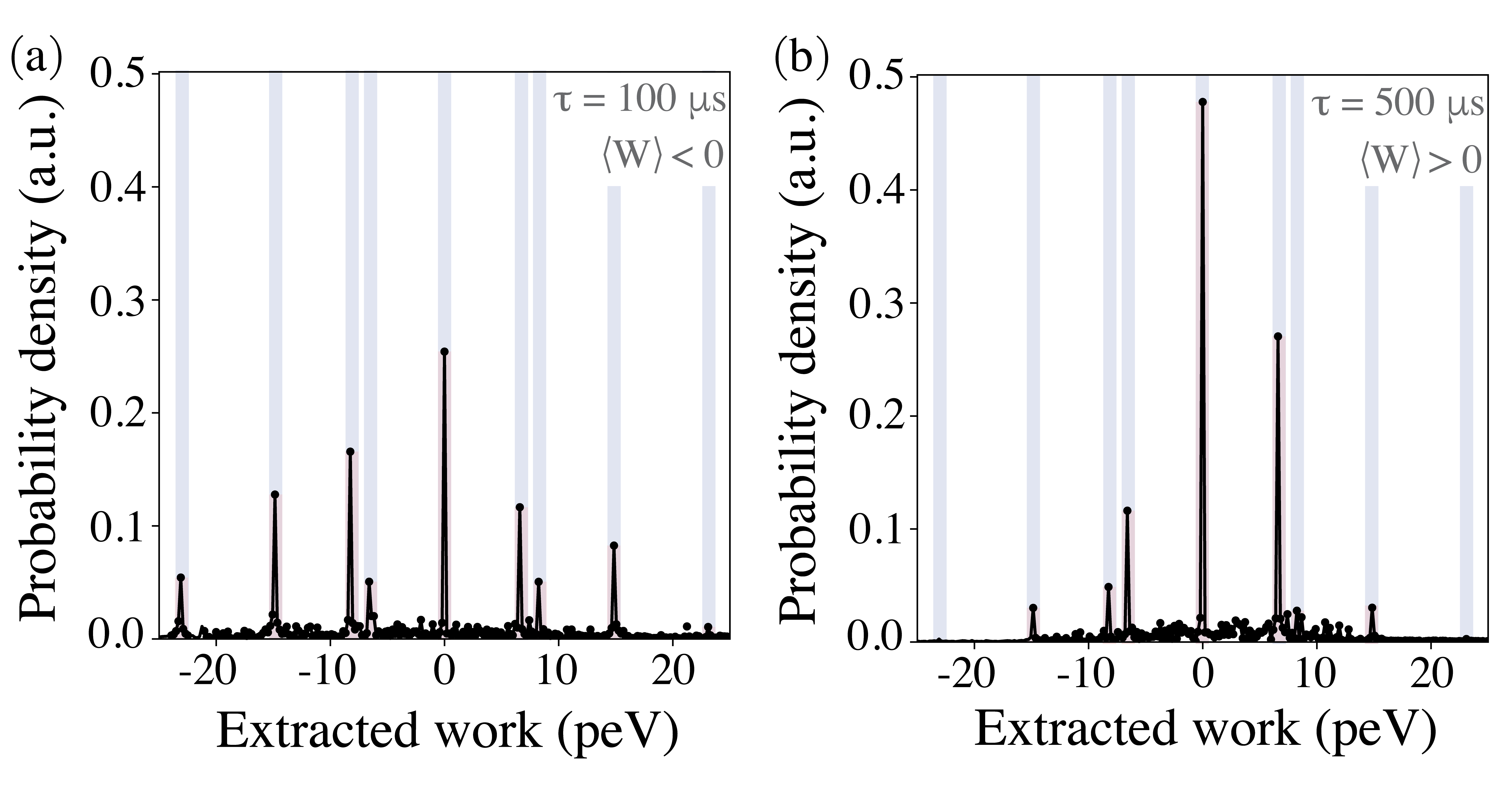} 
\par\end{centering}
\caption{Extracted work probability distribution of the quantum engine with
Hamiltonian driving time lengths: (a) $\tau=100\,\mu$s and (b) $\tau=500\,\mu$s.
Cold and hot source temperatures are set at $k_{B}T_{1}=\left(6.6\pm0.1\right)$~peV
and $k_{B}T_{2}^{B}=\left(40.5\pm3.7\right)$~peV, respectively.
The experimental data (points) is well fitted by a sum of nine Lorentzian
peaks (the full line) centered approximately at $0,$ $\pm h\left(\nu_{2}-\nu_{1}\right)$,
$\pm\nu_{1}$, $\pm\nu_{2}$, and $\pm h\left(\nu_{2}+\nu_{1}\right)$
(dashed columns), in agreement with the theoretical expectation (see
supplemental Fig.~S3 in \cite{SuppMat}). The error bars are smaller
than the symbols size and are not shown.}
\label{fig:WorkProbability} 
\end{figure}

Work extracted from (performed on) the $^{13}$C nuclear spin during
the energy gap expansion (compression) driving protocol is actually
a stochastic variable, described by a probability distribution \cite{Campisi2014,Campisi2015},
$P_{exp}(W)$ ($P_{comp}(W)$). The full thermalization with the hot
source allows us to write the work performed in each Hamiltonian driving
stroke of the cycle as independent variables. So, the net extracted
work from the engine is a convolution of the two marginal work probability
distributions, which can be assessed by the interferometric approach
\cite{Batalhao2014,Batalhao2015}. In the present experiment, the
characteristic function of the work probability distribution is measured.
In the energy gap expansion stroke, it is given by 
\begin{align}
\chi_{exp}\left(u\right) & =\text{tr}\left[\mathcal{U}_{\tau}e^{-iu\mathcal{H}_{\text{exp},0}^{\text{C}}}\rho_{0}^{\text{eq},1}\left(e^{-iu\mathcal{H}_{\text{exp},\tau}^{\text{C}}}\mathcal{U}_{\tau}\right)^{\dagger}\right]\nonumber \\
 & =\sum_{n,m=0}^{1}p_{n}^{0}p_{m\mid n}^{\tau}e^{iu\left(\epsilon_{m}^{\tau}-\epsilon_{n}^{0}\right)}\;,
\end{align}
where $p_{n}^{0}$ is the occupation probability of the $n$-th energy
level in the cold initial thermal state ($\rho_{0}^{\text{eq},1}$),
$p_{m|n}^{\tau}=\xi+\left(1-2\xi\right)\delta_{m,n}$ is the transition
probability between the Hamiltonian eigenstates induced by the time-dependent
quantum dynamics, $\epsilon_{m}^{\tau}$ and $\epsilon_{n}^{0}$ are
eigenenergies of the Hamiltonians $\mathcal{H}_{1}^{\text{C}}$ and
$\mathcal{H}_{2}^{\text{C}}$, respectively. Analogous expressions
hold for the compression stroke ($\chi_{comp}(u)$) \cite{SuppMat}.
The characteristic function for the engine net work is the product
of characteristic functions for both Hamiltonian driving protocols,
i.e. $\chi_{eng}(u)=\chi_{comp}(u)\chi_{exp}(u)$. Thus, the inverse
Fourier transform of the measured $\chi_{eng}(u)$ provides the work
probability distribution for the quantum engine as $P_{eng}(W)=\int du\chi_{eng}(u)e^{iuW}$
and the mean value of the extracted work can be obtained from the
statistics as $\left\langle W_{eng}\right\rangle =\int dWP_{eng}(W)W$.

\begin{figure*}
\begin{centering}
\includegraphics[scale=0.28]{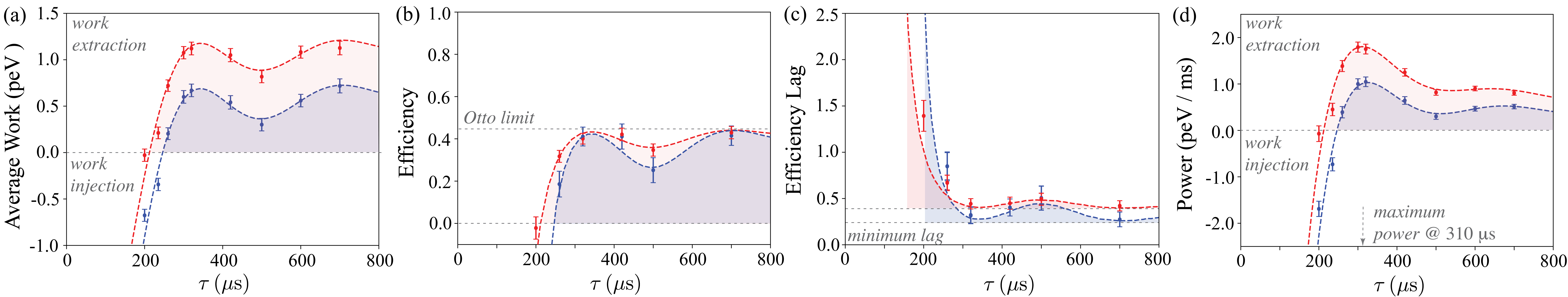} 
\par\end{centering}
\caption{Spin quantum engine figures of merit: (a) average extracted work,
(b) efficiency, (c) efficiency lag due to entropy production cf. Eq.~\eqref{efficiency2}
(the minimum lag is $\eta_{\text{Carnot}}-\eta_{\text{Otto}}$), and
(d) extracted power, as a function of the driving protocol time length
($\tau$). Points represent experimental data. The dashed lines are
based on theoretical predictions and numerical simulations. In all
experiments, the spin temperature of the cold source is set at $k_{B}T_{1}=\left(6.6\pm0.1\right)$~peV.
Data in blue and red correspond to implementations with the hot source
spin temperatures set at $k_{B}T_{2}^{A}=\left(21.5\pm0.4\right)$~peV
and $k_{B}T_{2}^{B}=\left(40.5\pm3.7\right)$~peV, respectively.}
\label{fig:FiguresOfMerit} 
\end{figure*}

We characterized the work distribution in different operation modes
of the spin engine, varying the driving time length ($\tau$) and
the hot source temperature, with representative results displayed
in Fig.~\ref{fig:WorkProbability}. The initial spin temperatures
of the $^{1}$H and $^{13}$C nuclei were verified by quantum state
tomography (QST) \cite{Oliveira2007}, which confirmed the Gibbs state
preparation. The spin temperature of the $^{13}$C cold initial state
is equivalent to $k_{B}T_{1}=\left(6.6\pm0.1\right)$~peV, while
the $^{1}$H was prepared in two hot states ($A$ and $B$) corresponding
to $k_{B}T_{2}^{A}=\left(21.5\pm0.4\right)$~peV and $k_{B}T_{2}^{B}=\left(40.5\pm3.7\right)$~peV.

There are nine observed peaks in Fig.~\ref{fig:WorkProbability}(a),
corresponding to the fastest implemented engine driving. A fit of
these experimental data allows us to determine the transition probability
$\xi$ that vary from ${\xi=0.02\pm0.02}$ (for $\tau=700\ensuremath{~}\mu$s)
to ${\xi=0.38\pm0.04}$ (for $\tau=100\ensuremath{~}\mu$s). We observe
that when the Hamiltonian driving is slower, as in Fig.~\ref{fig:WorkProbability}(b),
some of the work distribution peaks get decreased to the point of
being barely noticeable amid the noise (associated with the Fourier
analysis), since the dynamics is getting closer to the adiabatic one.
We also characterize the Hamiltonian driving protocol by means of
quantum process tomography \cite{Chuang1997,Nielsen2011} to certify
that it implements an almost unitary process \cite{SuppMat}.

The absorbed heat from the hot source ($^{1}$H nuclear spin) is also
a stochastic variable and its probability distribution, $\mathcal{P}(Q)$,
could be assessed by a two-time energy measurement scheme \cite{Talkner2007}.
However, in a full thermalization process, the measurement of energy
at the end of the process is uncorrelated with the measurement at
the start. Then, two QSTs are enough to provide a direct evaluation
of the heat probability distribution in the present implementation
\cite{SuppMat}. One of them is done at the end of the energy gap
expansion stroke (where the state is typically out-of-equilibrium),
while the other is done at the start of the energy gap compression
stroke (and thus should result in the hot thermal state). So the mean
heat from the hot source can be expressed as $\left\langle Q_{hot}\right\rangle =\int dQ\mathcal{P}(Q)Q$.

With the aforementioned data, we have fully characterized the quantum
heat engine. Its performance can be rated according to the average
work extracted per cycle, efficiency, efficiency lag, and the average
delivered power. These figures of merit are shown in Fig.~\ref{fig:FiguresOfMerit}(a)-\ref{fig:FiguresOfMerit}(d).
The work extraction regime requires a lower bound on the driving time
length, as can be seen in Fig.~\ref{fig:FiguresOfMerit}(a) and also
was anticipated by condition \eqref{transprob}. If the engine is
operated at a too-fast driving time length $\tau$ (smaller than $\approx200\,\mu\text{s}$
in this case), the entropy production is so large that it is not possible
to extract work. This entropy production decreases with a slower operation
rate, although not monotonically. The latter fact is a consequence
of the specific form of the Hamiltonian time modulation employed in
our implementation and does not generalize to other drivings.

Figure~\ref{fig:FiguresOfMerit}(b) illustrates that slower operation
leads to better efficiency. Nonetheless, the quantum engine irreversibility
can also be characterized by the efficiency lag \eqref{efficiency2}
measured by QST at different strokes. We observe a complete agreement
between the lag displayed in Fig.~\ref{fig:FiguresOfMerit}(c) and
the efficiency measured as the mean work and heat ratio {[}Fig.~\ref{fig:FiguresOfMerit}(b)
{]}. For the implemented quantum cycle, the main source of irreversibility
is the divergence (accounted by the relative entropy) of the state
achieved after the Hamiltonian driving protocols (expansion and compression)
and the reference (hot and cold) thermal states.

We are often interested in power, and a too-slow engine operation,
as an adiabatic dynamics, cannot deliver a fairly good amount of power.
Extracted power is maximized when the energy gap expansion (compression)
protocol takes about $310\,\mu\text{s}$ as can be noted in Fig.~\ref{fig:FiguresOfMerit}(d).
Quicker protocols are worse due to considerable entropy production
associated with energy level transitions during the dynamics {[}Fig.~\ref{fig:FiguresOfMerit}(c){]},
while slower driven protocols are also worse since they take more
time to deliver a similar amount of work {[}Fig.~\ref{fig:FiguresOfMerit}(a){]}.
The effective full thermalization with the hot source ($^{1}$H nucleus)
employed in our experiment {[}central part of the pulse sequence in
Fig.~\ref{fig:schematics-1}(b){]} lasts for about $7$ ms and it
takes the same time length in all operation modes of the spin engine.
In this fashion, we have opted to describe all results in terms of
the expansion and compression Hamiltonian driving time length $\tau$,
which is the finite-time feature in the present spin engine implementation.

We performed an experimental proof-of-concept of a quantum heat engine
based on a nuclear spin where the typical energy gaps, about $8.27$~peV,
are of the order of heat source energy, $k_{B}\left(T_{2}-T_{1}\right)$
($\approx15$~peV). The extracted work per cycle may be on the same
order of magnitude (few peV) depending on the driving protocol. At
maximum power ($\tau\approx310\ensuremath{~}\mu$s), the engine efficiency,
$\eta=42\pm6$\%, is very close to the Otto limit, $\eta_{Otto}=44$\%,
for the compression factor employed in the present implementation.
The power delivered by the quantum engine, in the finite-time operation
mode, is ultimately limited by quantum fluctuations (transitions among
the instantaneous energy eigenstates), which are also related to entropy
production \cite{Batalhao2015,Camati2016} leading to a ``quantum
friction'' \cite{Plastina2014,Feldmann2003}. Assessing the statistics
of energy fluctuations in the implemented engine, we fully characterize
its irreversibility and efficiency lag. The investigation of this
data can also allow the quantum engine optimization by choosing optimal
driving protocols.

The methods employed here to assess energy fluctuations and to characterize
irreversibility in the quantum engine are versatile and can be applied
to other experimental settings. The developed spin engine architecture
is a comprehensive platform for future investigations of thermodynamical
cycles at micro-scale, which would involve, for instance, non-equilibrium,
non-classical, and correlated heat sources, allowing the detailed
study of a plethora of effects in quantum thermodynamics \cite{Goold2016,Vinjanampathy2016}. 
\begin{acknowledgments}
\emph{Acknowledgments}. We thank E. Lutz, M. Paternostro, L. C. Céleri,
C. I. Henao, P. A. Camati, K. Micadei, and F. L. Semião for valuable
discussions. We acknowledge financial support from UFABC, CNPq, CAPES,
FAPERJ, and FAPESP. J.P.S.P. thanks support from Innovation, Science
and Economic Development Canada, the Government of Ontario, and CIFAR.
T.B.B. acknowledges support from National Research Foundation (Singapore),
Ministry of Education (Singapore), and United States Air Force Office
of Scientific Research (FA2386-15-1-4082). R.M.S. gratefully acknowledges
financial support from the Royal Society through the Newton Advanced
Fellowship scheme (Grant no. NA140436) and the technical support from
the Multiuser Experimental Facilities of UFABC. This research was
performed as part of the Brazilian National Institute of Science and
Technology for Quantum Information (INCT-IQ). 
\end{acknowledgments}

\global\long\def\thesection{S-\Roman{section}}
 \setcounter{section}{0} \global\long\def\thefigure{S\arabic{figure}}
 \setcounter{figure}{0} \global\long\def\theequation{S\arabic{equation}}
 \setcounter{equation}{0}\global\long\def\thetable{S\Roman{table}}
 \setcounter{table}{0}
 

\section*{Supplemental Material}

We provide here supplementary details about the experimental protocol
and data analysis.

\subsection*{Experimental setting and characterization of the initial state preparation}

The liquid sample comprises $50$~mg of $99$\% $^{13}$C-labeled
CHCl$_{3}$ (Chloroform) diluted in $0.7$~ml of 99.9\% deutered
Acetone-d6, in a flame sealed $5$~mm NMR tube. The experiments were
performed in a Varian $500$~MHz Spectrometer equipped with a double-resonance
probe-head. The sample is inserted at the center of a superconducting
magnet (immersed in liquid He inside a thermally shielded vessel)
within the radio frequency (rf) coil of the inner probe head. The
superconducting magnet produces a high intensity static magnetic field
in the longitudinal direction (which defines the $z$ axis). Magnetization
of the $^{1}$H and $^{13}$C nuclear spins (with Larmor frequencies
about $500$~MHz and $125$~MHz, respectively) can be controlled
by time-modulated rf-field pulses in the transverse ($x$ and $y$)
direction and longitudinal field gradient pulses.

Employing spatial average techniques, the $^{1}$H and $^{13}$C nuclei
are initially prepared in a pseudo-state equivalent to a tensor product,
$\rho_{0}^{eq,2}\otimes\rho_{0}^{eq,1}$, of thermal Gibbs states,
at spin temperatures $T_{2}$ (hot) and $T_{1}$ (cold), see also
the description of the thermal state initialization method in Ref.~\cite{Micadei2017}.
The initial pseudo-thermal state is certified by quantum state tomography
(QST) \cite{Oliveira2007}. The effective spin temperature of the
initial $^{1}$H ($^{13}$C) Gibbs state is related to the ground,
$p_{0}^{\text{H(C)}}$, and exited, $p_{1}^{\text{H(C)}}$, populations
as $k_{B}T_{2(1)}=h\nu_{2(1)}\left(\ln\frac{p_{0}^{\text{H(C)}}}{p_{1}^{\text{H(C)}}}\right)^{-1}$.
From the QST tomography data we verified the initial spin temperature
as displayed in Tab.~SI.

\begin{table}[h]
\begin{centering}
\caption{Population and spin temperatures of the Hydrogen and Carbon nuclei
initial states. The coherence (off-diagonal) elements of initial pseudo-thermal
states are null within the measurement error. }
\begin{tabular}{c|ccc}
\hline 
$^{1}$H nucleus  & $p_{0}^{\text{H}}$  & $p_{1}^{\text{H}}$  & $k_{B}T_{2}$ (peV)\tabularnewline
\hline 
option A  & $0.67\pm0.01$  & $0.33\pm0.01$  & $21.5\pm0.4$\tabularnewline
option B  & $0.60\pm0.01$  & $0.42\pm0.01$  & $40.5\pm3.7$\tabularnewline
\hline 
$^{13}$C nucleus  & $p_{0}^{\text{C}}$  & $p_{1}^{\text{C}}$  & $k_{B}T_{1}$ (peV)\tabularnewline
\hline 
 & $0.78\pm0.01$  & $0.22\pm0.01$  & $6.6\pm0.1$\tabularnewline
\hline 
\end{tabular}
\par\end{centering}
\label{tab:statetomography} 
\end{table}

\subsection*{Energy gap expansion and compression protocols}

The energy gap expansion and compression driven protocols are implemented
by a time-modulated rf field set at the Carbon Larmor frequency, in
order to produce the dynamics described by the time-dependent Hamiltonians
$\mathcal{H}_{exp}^{\text{C}}\left(t\right)$ and $\mathcal{H}_{comp}^{\text{C}}\left(t\right)$,
as defined in the main text. The time lengths of the driving processes
implemented (i.e. ${\tau=100}$,~$200$,~$235$,~$260$,~$300$,~$320$,~$420$,~$500$,~$600$,
and $700$~$\mu$s) are much shorter than the typical decoherence
time scale. Spin-lattice relaxation times for the Hydrogen and Carbon
nuclear spins are $\mathcal{T}_{1}^{\text{H}}=7.42$~s and $\mathcal{T}_{1}^{\text{C}}=11.31$~s,
respectively. Transverse relaxations were measured as $\mathcal{T}_{2}^{\text{*H}}=1.11$~s
and $\mathcal{T}_{2}^{*\text{C}}=0.30$~s. Besides the decoherence
phenomenon that has a small effect in the driving dynamics, experimental
imperfections and non-idealities prevent the realization of a perfectly
unitary evolutions associated with the driving Hamiltonians. In order
to verify the actually implemented protocol, we employed quantum process
tomography (QPT) \cite{Chuang1997,Nielsen2011}.

A general description of a quantum dynamics as a map $\mathcal{E}\left(\rho\right)$
acting on an initial density operator, can be done in terms of the
Choi-Jamiolkowski matrix, $\Upsilon$, through the relation 
\begin{equation}
\mathcal{E}\left(\rho\right)=\sum_{k,j=0}^{3}\Upsilon_{k,j}\Xi_{k}\rho\Xi_{j}^{\dagger},
\end{equation}
where $\Xi_{0}=i\mathbf{1}$ and $\left(\Xi_{1},\Xi_{2},\Xi_{3}\right)$
are the Pauli operators $\left(\sigma_{x},\sigma_{y},\sigma_{z}\right)$.
When written in this operator basis, any unital process (the one that
preserve the identity $\mathcal{E}\left(\mathbf{1}\right)=\mathbf{1}$)
corresponds to a real process matrix $\Upsilon$.

\begin{figure}
\begin{centering}
\includegraphics[width=0.45\textwidth]{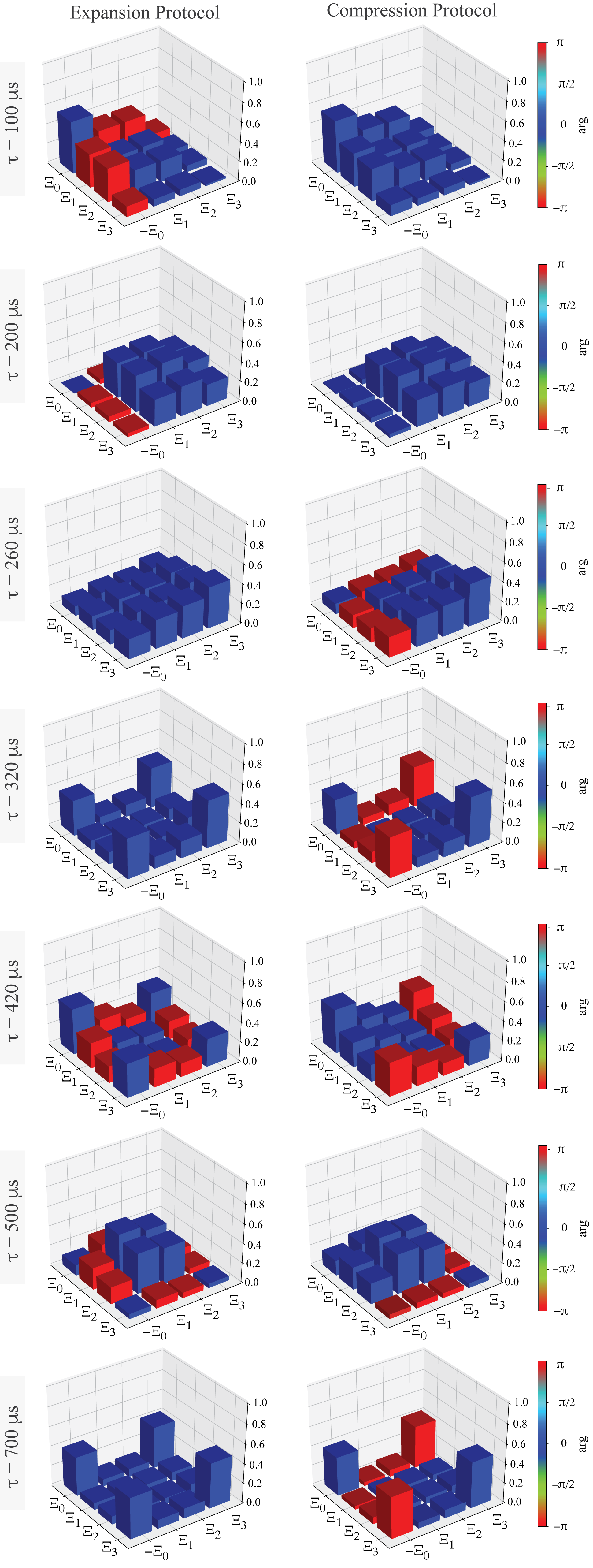} 
\par\end{centering}
\caption{Quantum process tomography experimental data. Process matrix representation
of the energy gap expansion and compression protocol. The height of
each element is proportional to its absolute value and the color depends
on its complex phase. Blue bars represent real positive values, while
red bars represent real negative elements. The imaginary part of each
element is approximately null within the experimental error.}
\label{fig:process_tomography} 
\end{figure}

We have prepared the $^{13}$C nuclear spin in a set of states of
a mutually unbiased basis (MUB) \cite{Chuang1997,Nielsen2011}, subjected
each of them to the expansion (compression) protocol to be characterized,
and finally reconstructed the output density matrix using QST. Numerical
post processing of the acquired data enables the final estimate of
the process matrix $\Upsilon$ elements. A summary of the results
is shown in Fig.~\ref{fig:process_tomography}. The height of each
element plotted in Fig.~\ref{fig:process_tomography} is proportional
to its absolute value, while the color depends on its complex phase.
We see that every element is either blue (representing real positive
elements) or red (representing real negative values), which is a visual
indication that the process is unital. In a more quantitative analysis,
typical values of the imaginary part are zero inside QST measurement
precision. We have also checked that the expansion and compression
processes, when applied in sequence, yields the identity process,
and found about $95$\% fidelity between a state and the result of
applying both processes to that state. It confirms that the compression
protocol is, in a good approximation, the time reversal version of
the expansion protocol.

\begin{figure}[h]
\begin{centering}
\includegraphics[width=0.48\textwidth]{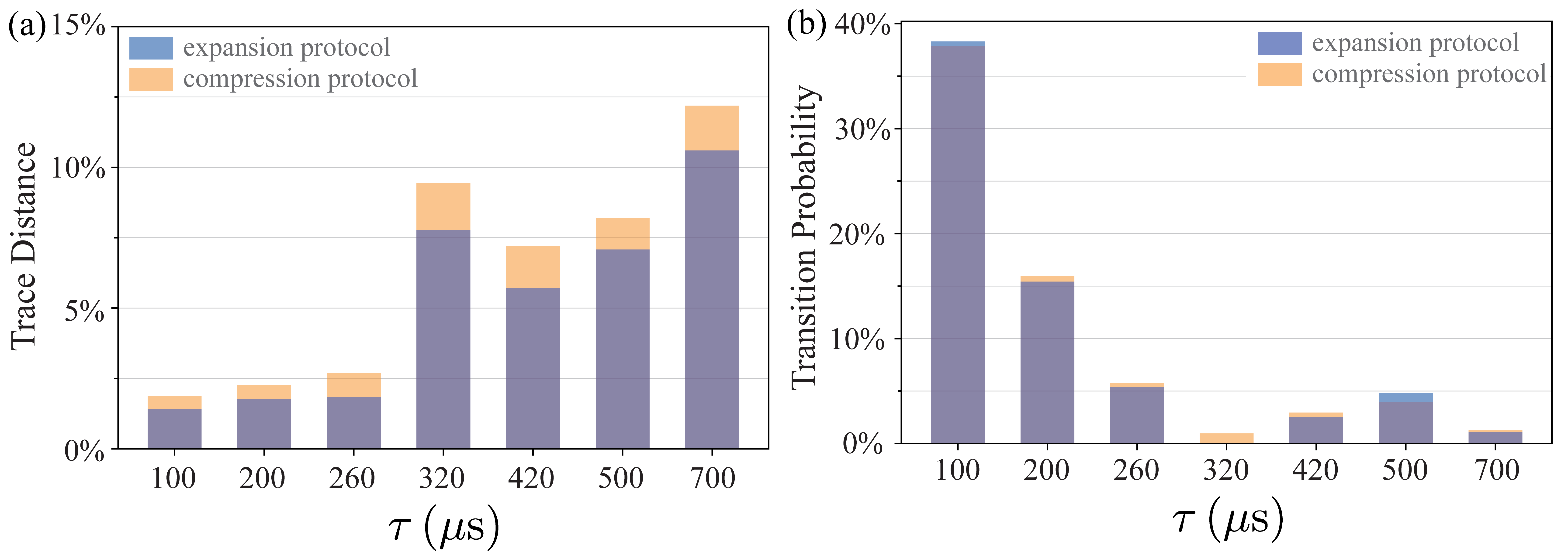} 
\par\end{centering}
\caption{Characterization of the implemented driving protocols. (a) Deviation
from a unitary evolution obtained from the QPT data quantifies by
the process trace distance ($\delta$). (b) Transition probability
among the instantaneous eigenstates ($\xi$), obtained from the work
probability distribution data, as a function of the Hamiltonian driving
protocol time length ($\tau$).}
\label{fig:protol-characterization} 
\end{figure}

From the eigenvalues and ``eigenoperators basis'' of the Choi-Jamiolkowski
matrix, we can check the closest unitary map to the experimentally
implemented process. The process trace distance, $\delta=\frac{1}{2}\text{tr}\left|\Upsilon^{exp}-\Upsilon^{id}\right|$,
between the map that describes the experimentally implemented driving
protocol ($\Upsilon^{exp}$) and the ideal unitary map ($\Upsilon^{id}$)
is plotted in Fig.~\ref{fig:protol-characterization}(a). It is associated
with the bias for distinguishing between the ideal and experimental
processes. We observe that the implemented driving protocols are almost
unitary for the fast-driving dynamics ($\delta<3$\%), as the time
length is increased the process trace distance also increases, but
it still relatively small (less than $12$\% for $\tau=700$~$\mu$s)
validating our approximation of the driving processes as quasi-unitary.
This small increasing in the process trace distance when we increase
$\tau$ is mainly due to spatial non-homogeneity of time-modulated
rf pulse and the transverse relaxation (which has more effect on long
dynamics).

\subsection*{Heating protocol}

From the local point of view of the Carbon nucleus the heating evolution,
has the effect of a linear non-unitary map $\mathcal{E}(\rho_{i})=\mathrm{Tr}_{i}\left(\mathcal{U}_{\tau}\rho_{A}^{0}\otimes\rho_{B}^{0}\mathcal{U}_{\tau}^{\dagger}\right)$
on the working substance, which can be represented as 
\begin{equation}
\mathcal{E}(\rho_{i})=\sum_{j=1}^{4}K_{j}\rho_{i}^{0}K_{j}^{\dagger}\label{eq:kraus-1}
\end{equation}
where $i=A,B$. The Kraus operators are given by 
\begin{align*}
K_{1} & =\sqrt{1-p}\begin{bmatrix}1 & 0\\
0 & 0
\end{bmatrix}, & K_{3} & =\sqrt{p}\begin{bmatrix}0 & 0\\
0 & 1
\end{bmatrix},\\
K_{2} & =\sqrt{1-p}\begin{bmatrix}0 & 1\\
0 & 0
\end{bmatrix}, & K_{4} & =\sqrt{p}\begin{bmatrix}0 & 0\\
-1 & 0
\end{bmatrix},
\end{align*}
with $p$ being the population of the excited state in the Hydrogen
nucleus.

This map is equivalent to the \textit{generalized amplitude damping}
which is the Kraus map for the thermalization (in this case full thermalization)
of a single spin-$1/2$ system. Therefore, from the local point of
view of the working substance, the interaction with the \textquotedbl hot
spin system\textquotedbl{} in the experiment is indistinguishable
from a thermalization map. This plays the role of an effective thermalization
in our proof of concept experiment.

\subsection*{Control rotations perform no work in the engine protocol}

Consider a single rotation (represented by the red and blue circles
in the pulse sequence description) implemented by a hard square rf
pulse on resonance with the nuclear spin A. The Hamiltonian of the
system, in the rotation frame with the Larmor frequency of the nuclear
spin A, can be effectively described as
\begin{equation}
\mathcal{H}=u(t)\mathcal{V}_{\text{A}},
\end{equation}
where $u(t)=\theta(t)-\theta(t-t')$ can very well modeled as the
sum of two (unity) Heaviside functions ($\theta(x)$) such that $u(t)$
is $1$ if $0<t<t'$ and $0$ if $t<0$ or $t>0$ and $\mathcal{V}_{\text{A}}$
is the potential generated by the transverse rf field. The mean work
performed in the process of turning on and off the rf pulse between
the two spin systems can be unambiguously defined as 
\begin{equation}
\left\langle W\right\rangle =\int_{-\infty}^{\infty}dt\left\langle \frac{\partial\mathcal{H}}{\partial t}\right\rangle _{t}
\end{equation}
Since $\dot{u}(t)=\delta(t)-\delta(t-t')$ (where $\delta(x)$ is
the Dirac delta function), it follows that 
\begin{equation}
\left\langle W\right\rangle =\langle\mathcal{V}_{\text{A}}\rangle_{0}-\langle\mathcal{V}_{\text{A}}\rangle_{t'}=0.
\end{equation}
Hence, no work is performed when the transient time for turning on
and off the time-independent rf rotation pulse is sufficiently small
to be modelled as (unity) Heaviside functions, which is precisely
the case in our experiment.

\subsection*{Extracted work statistics}

\begin{figure*}[th]
\begin{centering}
\includegraphics[scale=0.17]{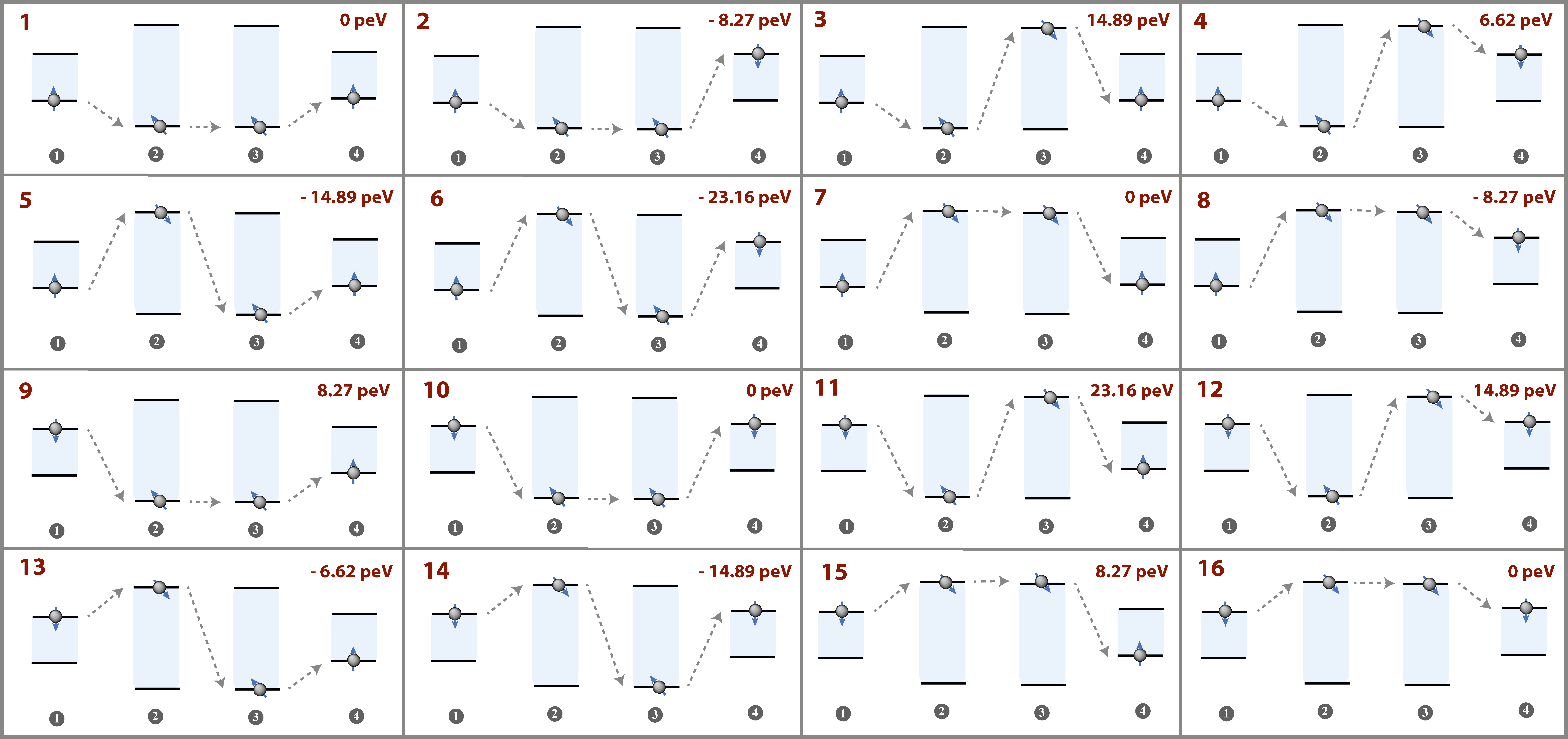} 
\par\end{centering}
\caption{Transitions between the instantaneous energy eigenstates of the spin-1/2
working medium performing the quantum Otto cycle. There are sixteen
possible energy-transition combinations in the spin engine. Variation
in the internal energy of each transition history is displayed on
the top right side of each inner box. The implemented four-stroke
cycle is illustrated as follows: (1) $^{13}$C nucleus starts in the
cold thermal state; (2) it is driven by a time-modulated rf field
that expand the energy gap resulting in a out-of-equilibrium state;
(3) the working medium thermalizes with the hot source; (4) it is
driven by the time reversal protocol that compress the energy gap.}
\label{fig:transitions-supp} 
\end{figure*}

The extracted work from (work performed on) the $^{13}$C nuclear
spin during the Otto cycle is a stochastic variable with a probability
distribution written in terms of a characteristic function as

\begin{align}
P_{eng}(W) & =\int du\chi_{eng}(u)e^{iuW}.
\end{align}
The work characteristic function of the spin engine performing the
quantum Otto cycle (with full thermalization) can be written as

\begin{align}
\chi_{eng}(u) & =\chi_{comp}(u)\chi_{exp}(u)\nonumber \\
 & =\sum_{n,m,k,j=0}^{1}p_{n}^{0}p_{m\mid n}^{\tau}q_{k}^{0}q_{j\mid k}^{\tau}e^{iu\left(\epsilon_{m}^{\tau}-\epsilon_{n}^{0}+\epsilon_{j}^{0}-\epsilon_{k}^{\tau}\right)},
\end{align}
where $p_{n}^{0}$ is the occupation probability of the $n$-th energy
level in the initial cold thermal state ($\rho_{0}^{\text{eq},1}$),
$p_{m|n}^{\tau}=\left|\left\langle m^{(\tau)}\left|\mathcal{U}_{\tau}\right|n^{(0)}\right\rangle \right|^{2}$
is the transition probability between the instantaneous eigenstates
$\left|n^{(0)}\right\rangle $ ($\left|m^{(\tau)}\right\rangle $)
of the driving Hamiltonian $\mathcal{H}_{exp}^{\text{C}}\left(0\right)$
($\mathcal{H}_{exp}^{\text{C}}\left(\tau\right)$), $q_{k}^{0}$ is
the occupation probability of the $k$-th energy level in the equilibrium
state ($\rho_{0}^{\text{eq},2}$) after the thermalization with the
hot source, $q_{m|n}^{\tau}=\left|\left\langle j^{(\tau)}\left|\mathcal{V}_{\tau}\right|k^{(0)}\right\rangle \right|^{2}$
is transition probability between the instantaneous eigenstates $\left|k^{(0)}\right\rangle $
($\left|j^{(\tau)}\right\rangle $) of the driving Hamiltonian $\mathcal{H}_{comp}^{\text{C}}\left(0\right)$
($\mathcal{H}_{comp}^{\text{C}}\left(\tau\right)$), $\epsilon_{m}^{\tau}$
and $\epsilon_{n}^{0}$ are eigenvalues of the Hamiltonians $\mathcal{H}_{2}^{\text{C}}$
and $\mathcal{H}_{1}^{\text{C}}$ (defined in the main text), respectively.

The work characteristic function of the spin engine can be rewritten
as

\begin{align}
\chi_{eng}(u)= & \text{tr}\left[\mathcal{U}_{\tau}e^{-iu\mathcal{H}_{\text{exp},0}^{\text{C}}}\rho_{0}^{\text{eq},1}\left(e^{-iu\mathcal{H}_{\text{exp},\tau}^{\text{C}}}\mathcal{U}_{\tau}\right)^{\dagger}\right]\nonumber \\
 & \times\text{tr}\left[\mathcal{V}_{\tau}e^{-iu\mathcal{H}_{\text{comp},0}^{\text{C}}}\rho_{0}^{\text{eq},2}\left(e^{-iu\mathcal{H}_{\text{comp},\tau}^{\text{C}}}\mathcal{V}_{\tau}\right)^{\dagger}\right].\label{eq:characteristic-supp}
\end{align}
Each trace (associated with the expansion and compression drivings)
in the product of rhs of Eq.~\eqref{eq:characteristic-supp} was
measured by an adaptation of the interferometric protocol described
in Refs.~\cite{Batalhao2014,Batalhao2015}.

\begin{figure}[h]
\begin{centering}
\includegraphics[width=0.48\textwidth]{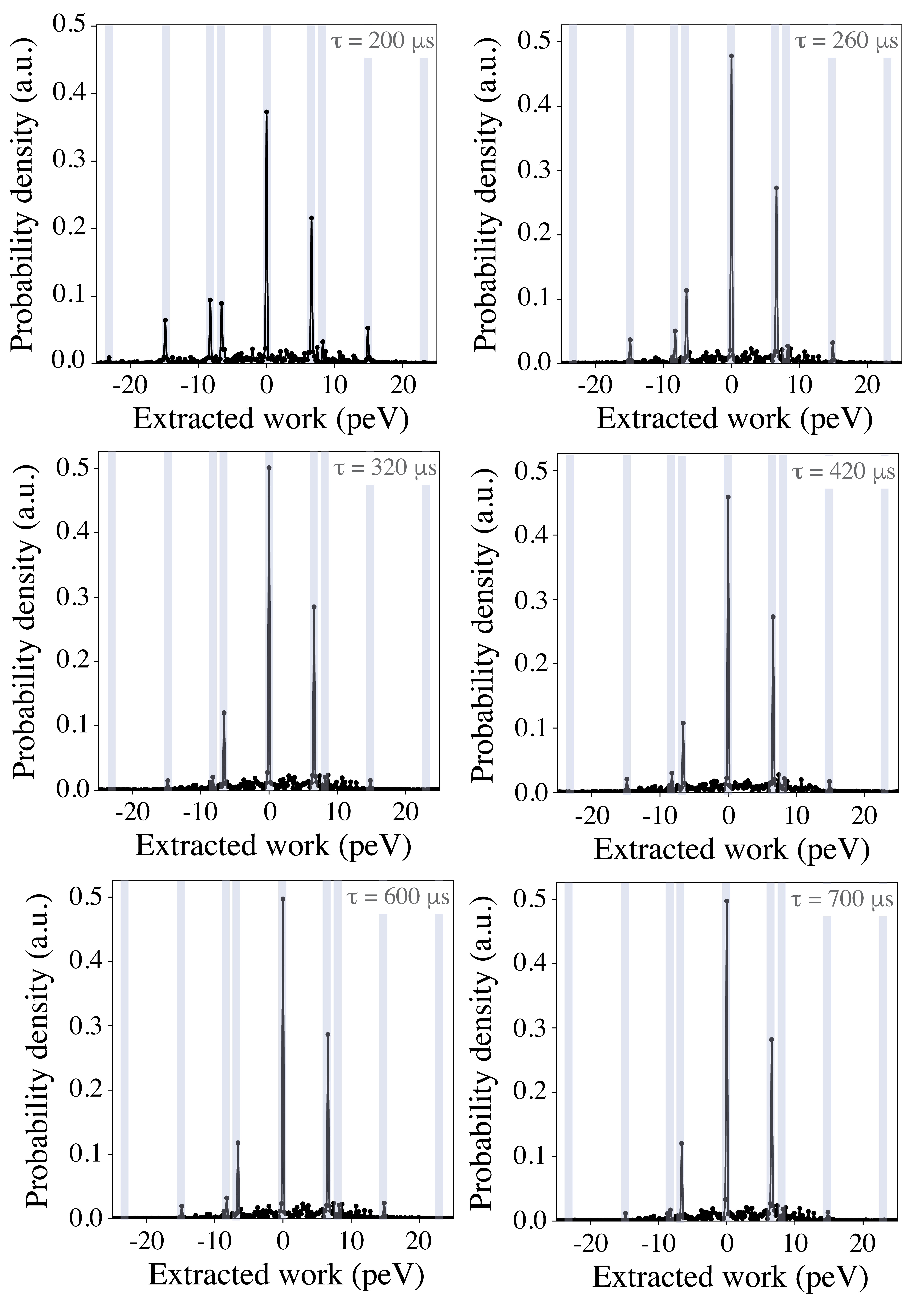} 
\par\end{centering}
\caption{Extracted work probability distribution of the quantum engine with
driving time lengths: ${\tau=200}$,~$260$,~$320$,~$420$,~$600$,
and $700$~$\mu$s. Cold and hot source temperatures are set at $k_{B}T_{1}=\left(6.6\pm0.1\right)$~peV
and $k_{B}T_{2}^{B}=\left(40.5\pm3.7\right)$~peV, respectively.
The experimental data (points) is well fitted by a sum of nine Lorentzian
peaks (the full line) centered approximately at $0,$ $\pm6.62$,
$\pm8.27$, $\pm14.89$, and $\pm23.16$~peV (dashed columns), in
agreement with the theoretical expectation (see Fig.~\ref{fig:transitions-supp}).
The error bars are smaller than the symbols size and are not shown.}
\label{fig:work-dist-supp} 
\end{figure}

As a consequence of the finite-time dynamics produced by the driving
protocol and the thermalization process, transitions between the instantaneous
energy eigenstates of the working medium may occur. Figure~\ref{fig:transitions-supp}
illustrates the sixteen possible energy-transition combinations of
the spin engine performing the Otto cycle. One or more histories of
Fig.~\ref{fig:transitions-supp} are associated with each peak in
the extracted work probability distribution plotted in Fig.~2 of
the main text and in Fig.~\ref{fig:work-dist-supp}. For the sake
of completeness, we plot in Fig.~\ref{fig:work-dist-supp} the work
probability distribution for driving time lengths not displayed in
the main text. From the experimental statistics of work, we determined
the transition probability among the instantaneous eigenstates ($\xi$)
as shown in Fig.~\ref{fig:protol-characterization}(b). As expected
for a fast Hamiltonian driving, we have a big transition probability
(about 38\% for $\tau=100$~$\mu$s). As the driving time length
increases, the transition probability decreases readily (reaching
about 2\% for $\tau=700$~$\mu$s), since we are getting close to
an adiabatic evolution.

\subsection*{Heat flow from the hot source}

\begin{figure}[h]
\begin{centering}
\includegraphics[width=0.4\textwidth]{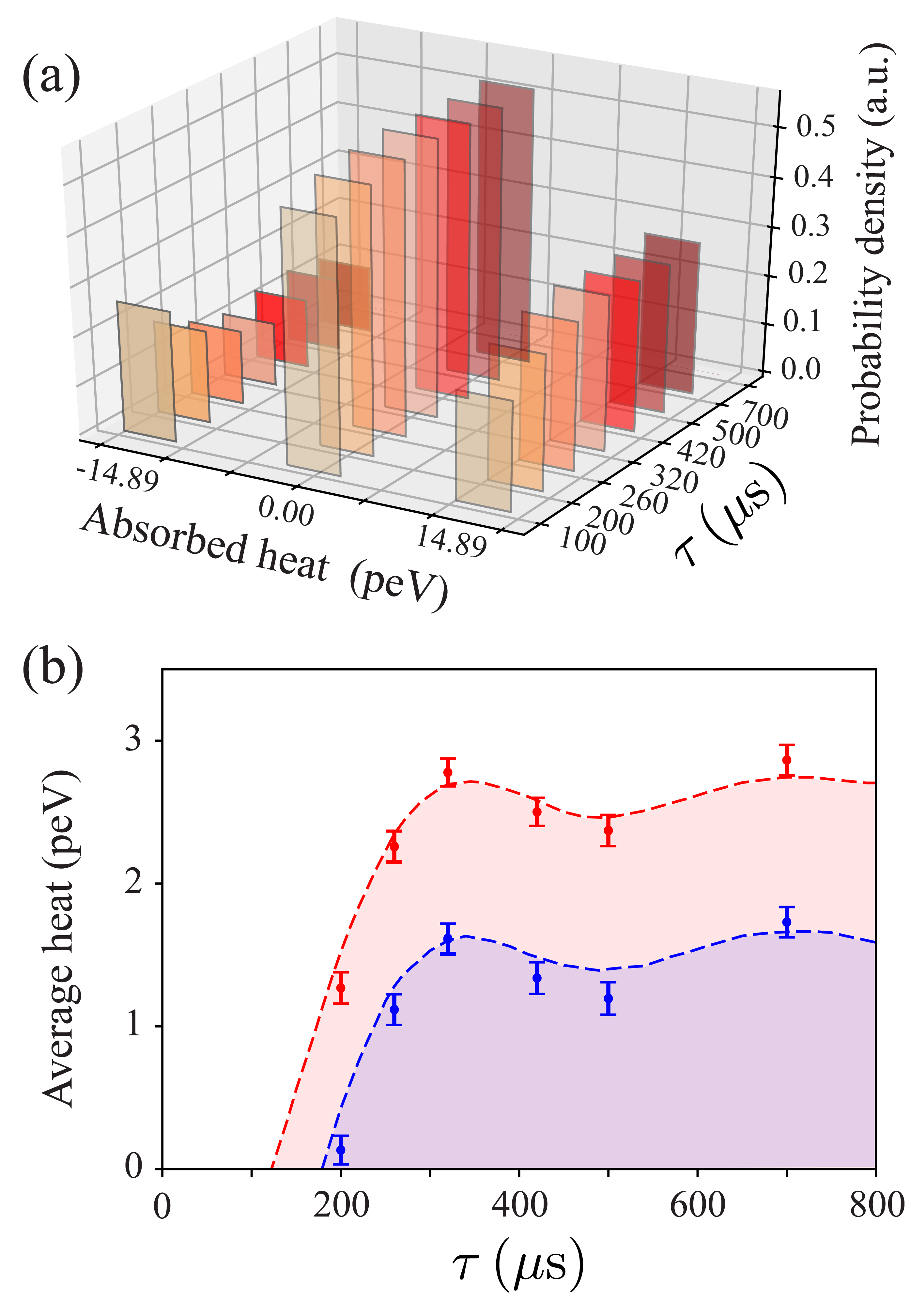} 
\par\end{centering}
\caption{Absorbed heat from the hot source by the spin quantum engine. (a)
Heat probability distribution, $\mathcal{P}(Q),$ with different Hamiltonian
driving time lengths $\tau$. Cold and hot source temperatures are
set at $k_{B}T_{1}=\left(6.6\pm0.1\right)$~peV and $k_{B}T_{2}^{B}=\left(40.5\pm3.7\right)$~peV,
respectively. The error bars are smaller than the size of the symbols
and are not shown. (b) Average heat from the hot source, $\left\langle Q_{hot}\right\rangle $
, as a function of $\tau$. Points represent experimental data. The
dashed lines are based on theoretical predictions and numerical simulations.
In all experiments, the spin temperature of the cold source is set
at $k_{B}T_{1}=\left(6.6\pm0.1\right)$~peV. Data in blue and red
correspond to implementations with the hot source spin temperatures
set at $k_{B}T_{2}^{A}=\left(21.5\pm0.4\right)$~peV and $k_{B}T_{2}^{B}=\left(40.5\pm3.7\right)$~peV,
respectively.}
\label{fig:heat-distrb} 
\end{figure}

The heat exchanged between the working medium ($^{13}$C nuclear spin)
and the hot source ($^{1}$H nuclear spin) in the heating stroke is
also a stochastic variable with a heat characteristic function giving
by

\begin{align}
\chi_{Q_{hot}}(u) & =\sum_{m,k=0}^{1}\left(\sum_{n=0}^{1}p_{n}^{0}p_{m\mid n}^{\tau}\right)q_{k}^{0}e^{iu\left(\epsilon_{m}^{\tau}-\epsilon_{k}^{\tau}\right)}.\label{eq:characteristic-heat-supp}
\end{align}
We note that the expression \eqref{eq:characteristic-heat-supp} does
not depends explicitly on the energy level transitions ($p_{m\mid n}^{\tau}$)
due to the expansion driving protocol. It only depends on the occupation
probability of the energy levels of the Hamiltonian $\mathcal{H}_{2}^{\text{C}}$
just after the expansion driving protocol, $s_{m}=\sum_{n=0}^{1}p_{n}^{0}p_{m\mid n}^{\tau}$,
that can be obtained through a QST performed before the heating stroke.
The occupation probability of the $k$-th energy level after the thermalization
with the hot source, $q_{k}^{0}$, can also be obtained by QST. The
marginal probability distribution for the heat from the hot source
can be written as

\begin{align}
\mathcal{P}(Q) & =\int du\chi_{Q_{hot}}(u)e^{-iuQ},\nonumber \\
 & =\sum_{m,k=0}^{1}\left(\sum_{n=0}^{1}p_{n}^{0}p_{m\mid n}^{\tau}\right)q_{k}^{0}\delta\left(\epsilon_{k}^{\tau}-\epsilon_{m}^{\tau}-Q\right).
\end{align}

The mean value of the absorbed heat from the hot source, $\left\langle Q_{hot}\right\rangle =\text{tr}\left[\mathcal{H}_{\text{exp},\tau}^{\text{C}}\left(\rho_{0}^{\text{eq},2}-\mathcal{U}_{\tau}\rho_{0}^{\text{eq},1}\mathcal{U}_{\tau}^{\dagger}\right)\right]$,
can also be acquired from the direct observation of the variation
of the $^{13}$C nucleus magnetization before and after the heating
process. Moreover, alternatively as another verification for the heat
flow measurement, data from QST and QPT can be combined to reconstruct
the working medium state at the beginning of each stroke in the Otto
cycle. The data obtained from different strategies to determine the
heat probability distribution is in good agreement.

Statistics of heat flow is gathered from the reconstructed density
matrix and it is displayed in Fig.~\ref{fig:heat-distrb}(a) for
different Hamiltonian driving time lengths $\tau$. The amount of
heat exchanged with the hot source depends on the final state of the
$^{13}$C nucleus reached after the expansion driving protocol. In
a fast energy expansion driving ($\tau=100$~$\mu$s), the final
state corresponds to a non-equilibrium state with not null coherences
(non-diagonal elements) in the Hamiltonian basis turning the heat
absorption less effective. On the other hand, in a slow energy expansion
driving ($\tau=700$~$\mu$s), the final state also corresponds to
a non-equilibrium state, but the coherences elements in the Hamiltonian
basis are much smaller (almost null). In fact, in the later case the
state after energy expansion driving is very close to the state obtained
in an adiabatic dynamics. The mean value of the absorbed heat $\left\langle Q_{hot}\right\rangle $
as a function of the Hamiltonian driving protocol time length ($\tau$)
is shown in Fig.~\ref{fig:heat-distrb}(b). The results obtained
for the mean value of the absorbed heat $\left\langle Q_{hot}\right\rangle $
are in complete agreement with the one obtained from the direct observation
of the $^{13}$C nucleus magnetization before and after the heating
stroke.

\subsection*{Efficiency Lag (Eq.~(5) of the main text)}

In this section we will outline a demonstration of the expression
for the Efficiency Lag introduced in Eq.~(5) of the main text. Let
us start with the sum of the entropy production in the expansion and
compression driving protocols of the quantum Otto cycle, which can
be expressed as the sum of two relative entropies that reads 
\begin{align}
\Sigma_{drive} & =\mathcal{S}\left(\left.\rho_{\tau}^{1}\right\Vert \rho_{0}^{eq,2}\right)+\mathcal{S}\left(\left.\rho_{\tau}^{2}\right\Vert \rho_{0}^{eq,1}\right)\nonumber \\
 & =\text{tr}\left(\rho_{\tau}^{1}\ln\rho_{\tau}^{1}\right)-\text{tr}\left(\rho_{\tau}^{1}\ln\rho_{0}^{eq,2}\right)\nonumber \\
 & \:\:\:\:+\text{tr}\left(\rho_{\tau}^{2}\ln\rho_{\tau}^{2}\right)-\text{tr}\left(\rho_{\tau}^{2}\ln\rho_{0}^{eq,1}\right),
\end{align}
where $\rho_{\tau}^{1}=\mathcal{U}_{\tau}\rho_{0}^{eq,1}\mathcal{U}_{\tau}^{\dagger}$
is the final non-equilibrium state after the expansion protocol and
$\rho_{\tau}^{2}=\mathcal{V}_{\tau}\rho_{0}^{eq,2}\mathcal{V}_{\tau}^{\dagger}$
is the final non-equilibrium state after the compression protocol.
The von Neumann entropy is invariant under the unitary driving protocol,
so it implies that $\text{tr}\left(\rho_{\tau}^{\alpha}\ln\rho_{\tau}^{\alpha}\right)=\text{tr}\left(\rho_{0}^{\alpha}\ln\rho_{0}^{\alpha}\right)$.
Using this fact and recalling the form of the initial pseudo-thermal
state, $\rho_{0}^{\text{eq},\alpha}=\left.e^{-\beta_{\alpha}\mathcal{H}_{\alpha}^{\text{C}}}\right/Z_{\alpha}$,
where $\alpha=1,2$, we obtain 
\begin{align}
\Sigma_{drive} & =-\beta_{1}\text{tr}\left[\left(\rho_{0}^{eq,1}-\rho_{\tau}^{2}\right)\mathcal{H}_{1}^{\text{C}}\right]\nonumber \\
 & \:\:\:\:-\beta_{2}\text{tr}\left[\left(\rho_{0}^{eq,2}-\rho_{\tau}^{1}\right)\mathcal{H}_{2}^{\text{C}}\right].
\end{align}
The last equation can be simplified to $\Sigma_{drive}=-\beta_{1}\left\langle Q_{cold}\right\rangle -\beta_{2}\left\langle Q_{hot}\right\rangle $,
where $\left\langle Q_{cold}\right\rangle =\text{tr}\left[\left(\rho_{0}^{eq,1}-\rho_{\tau}^{2}\right)\mathcal{H}_{1}^{\text{C}}\right]$
and $\left\langle Q_{hot}\right\rangle =\text{tr}\left[\left(\rho_{0}^{eq,2}-\rho_{\tau}^{1}\right)\mathcal{H}_{2}^{\text{C}}\right]$.
Employing the First Law of thermodynamics, $\left\langle Q_{cold}\right\rangle +\left\langle Q_{hot}\right\rangle -\left\langle W_{eng}\right\rangle =0$,
we can eliminate the heat from the cold source and write $\Sigma_{drive}=-\beta_{1}\left\langle W_{eng}\right\rangle +\left(\beta_{1}-\beta_{2}\right)\left\langle Q_{hot}\right\rangle $,
which lead us to the relation 
\begin{equation}
\frac{\left\langle W_{eng}\right\rangle }{\left\langle Q_{hot}\right\rangle }=\left(1-\frac{T_{1}}{T_{2}}\right)-\frac{\mathcal{S}\left(\left.\rho_{\tau}^{1}\right\Vert \rho_{0}^{eq,2}\right)+\mathcal{S}\left(\left.\rho_{\tau}^{2}\right\Vert \rho_{0}^{eq,1}\right)}{\beta_{1}\left\langle Q_{hot}\right\rangle }.\label{eq:effrelation}
\end{equation}

The lhs of Eq.~\eqref{eq:effrelation} can be identified as the efficiency
of the quantum heat engine ($\eta$). The first and the second terms
of the rhs are the Carnot limit ($\eta_{\text{Carnot}}$) and the
efficiency lag ($\mathcal{L}$), respectively. The above development
demonstrates the definition introduced in Eq.~(5) of the main text
for the quantum Otto cycle. The efficiency lag, $\mathcal{L}$, describes
the quantum engine irreversibility in terms of its microscopic state
dynamics.

\subsection*{Error analysis}

The most relevant sources of error in the experiments are small non-homogeneities
of the transverse rf field, non-idealities in its time modulation,
and non-idealities in the longitudinal field gradient. To estimate
the error propagation, we have employed a Monte Carlo method, to sample
deviations of the QST and magnetization data with a Gaussian distribution
having widths determined by the corresponding variances. The standard
deviation of the distribution of values for the relevant quantities
is estimated from this sampling. The variances of the tomographic
data are obtained by preparing the same state one hundred times, taking
the full state tomography and comparing it with the theoretical expectation.
These variances include random and systematic errors in both state
preparation and data acquisition by QST. The error in each element
of the density matrix estimated from this analysis is about 1\%. All
control parameters in the experimental implementation, such as pulse
intensity, phase, and its time length, are optimized in order to minimize
errors. 

\end{document}